\documentclass[aps,preprint,floats,epsf,epsfig,nofootinbib,letter]{revtex4-1}
\usepackage{epsfig}

\def\bea{\begin{eqnarray}}
\def\eea{\end{eqnarray}}

\def\lsim{ {\ \lower-1.2pt\vbox{\hbox{\rlap{$<$}\lower5pt\vbox{\hbox{$\sim$}
}}}\ } }
\def\gsim{ {\ \lower-1.2pt\vbox{\hbox{\rlap{$>$}\lower5pt\vbox{\hbox{$\sim$}
}}}\ } }

\begin{document}
\begin{flushright}
{\small
CYCU-HEP-12-04
}
\end{flushright}

\title{\Large \bf Galactic Dark Matter in the Phantom Field}

\author{\vspace*{0.3cm} \sc Ming-Hsun Li}
\author{ \sc Kwei-Chou Yang}

\bigskip
\affiliation{\vspace*{0.3cm}  Department of Physics,
Chung Yuan Christian University, Chung-Li 320, Taiwan \vspace*{2.3cm} }


\begin{abstract}{\small

 We investigate the possibility that the galactic dark matter exists in an scenario
where the phantom field is responsible for the dark energy.  We obtain the statically and spherically approximate solution for this kind of the galaxy system with a  supermassive black hole at its center.  The solution of the metric functions is satisfied with $g_{tt} = - g_{rr}^{-1}$. Constrained by the observation of the rotational stars moving in circular orbits with nearly constant tangential speed in a spiral galaxy, the background of the phantom field which is spatially inhomogeneous has an exponential potential. To avoid the well-known quantum instability of the vacuum at high frequencies, the phantom field defined in an effective theory is valid only at low energies.  
 Under this assumption, we further investigate the following properties. The absorption cross section of the low-energy $S$-wave excitations of the phantom field into the central black hole is shown to be the horizontal area of the central black hole.  Because the infalling phantom particles have a total negative energy,  the accretion of the phantom energy is related to the decrease of the black hole mass which  is estimated to be much less than a solar mass in the lifetime of the Universe. Using a simple model with the cold dark matter very weakly coupled to the "{\it low-frequency}" phantom particles which are generated from the background,  we show that these two densities can be quasi-stable in the galaxy.
}
\end{abstract}
\small
\maketitle

\section{Introduction}

The recent experimental data have shown that the current Universe is undergoing a phase of accelerated expansion \cite{Perlmutter:1998np,Riess}.
Considering the universe filled with a barotropic perfect fluid which corresponds to the dark energy component, its equation of state $w<-1/3$ is required for cosmic acceleration, where $w=p/\rho$ with $\rho$ and $p$ being the density and pressure, respectively. Recently observations suggest that  the equation of state lies in a narrow strip around $w =-1$ \cite{Komatsu:2010fb,Alam:2003fg}, where  $w =-1$ corresponds to a cosmological constant $\Lambda$ and $w <-1$ is allowed. A specific form of the dark energy corresponding to the phantom field was proposed to realize the possibility  of late-time acceleration with $w <-1$ \cite{Caldwell:1999ew,Caldwell:2003vq}, while the quitom model has $w$ crossing $-1$ \cite{Cai:2009zp}. The peculiar property of the phantom dark energy is the violation of the dominant energy condition (DEC), so that the energy density and curvature may grow to infinity in a finite time, which is referred to a Big Rip singularity \cite{Caldwell:2003vq,Nesseris:2004uj}.

Observations related to the cosmic microwave background (CMB)  and the large-scale structure (LSS) support that the Universe is very close to spatially flat geometry \cite{spergel:2003,Komatsu:2010fb}. At the present time our universe is dominated by dark energy with the fraction $\sim 72\%$. The most  accessible component of the Universe is baryonic matter which amounts to only 4.6\%.  The main remaining part that is non-baryonic and non-luminous is  believed to be the so-called dark matter responsible for $\sim  23\%$.  The weakly interacting massive particles (WIMPs) are considered one of the main candidates for cold dark matter which is dust-like with equation of state $w\simeq 0$. A dark matter halo, demonstrated by its gravitational effect on a spiral galaxy's rotation curve, dominates the mass of a galaxy and can stretch up to be larger than 50 kpc from the center of a galaxy.

The presence of the interaction between the dark matter and dark energy field may modify the distribution of dark matter in a galaxy.  A homogeneous scalar field with negative kinetic energy  has been used to model the phantom energy (with $w<-1$) and then to study  the cosmological evolution \cite{Carroll:2003st,Copeland:2006wr,Guo:2004xx}, where the scalar field varies only with time but does not change spatially. However,  for a galaxy due to the gravitational instabilities the phantom field may not be spatially homogeneous. Some works related to the inhomogeneous dark energy properties have been completed \cite{Sushkov:2005kj,Zaslavskii:2005fs,Kuhfittig:2006xj,Lobo:2005us,Lobo:2005yv,Lobo:2005uf,DeBenedictis:2008qm,Dzhunushaliev:2008bq,Yazadjiev:2011sm}. In this paper, we are interested in the static solution of the Einstein equations that can describe the dark matter halo with the existence of the supermassive black hole at its center and the background of the spatially inhomogeneous  phantom field.  We adopt the standard assumption that the dark matter halo consists of the WIMPs with $w\simeq 0$. We find that an approximate solution of the metric exists for describing this galactic halo scenario when we take the metric functions to be satisfied with $g_{tt} =- g_{rr}^{-1}$.  In general, several static and spherically  symmetric exact solutions of Einstein equations are obtained in terms of the parameter $\epsilon$, defined by $T^\theta_{\ \theta} = T^r_{\ r} (1-\epsilon)$, for which some related works can be found in Refs. \cite{Salgado:2003ub,Dymnikova:2001fb,Giambo:2002wr,Kiselev:2002dx}. We also obtain the approximate solution for the spatially inhomogeneous  phantom field in a  galaxy. Our result indicates that the corresponding exponential potential of the phantom field is relevant to the stage of  structure formation.

We further study the stability of the space-time structure for the galaxy. Since the  phantom dark energy field can cause the quantum instability of the vacuum at high frequencies, we therefore  treat it  as  an effective theory valid at low energy, i.e., we add a cutoff in the momentum integral \cite{Carroll:2003st}.    The stringent limit for the cutoff is $\Lambda \lesssim 3$ MeV, which was obtained from the diffuse gamma ray background \cite{Cline:2003gs}.
We compute the accretion rate of  the phantom particles into the black hole.  Since the infalling phantom particles have a total negative energy, the black hole mass diminishes in the process. On the other hand, due to the conservation of the angular momentum of the individual WIMPs and their tiny interaction rate, the capture cross section for dark matter particles by the supermassive black hole is sufficiently small. Therefore, the dark matter that we consider here is quite stable.

This paper is organized as follows. In Sec. \ref{sec:exact}, we consider the pressless  massive dark matter (WIMP) directly or indirectly interacting with a phantom field which is associated with the acceleration of the Universe. We solve the Einstein equations with the condition $g_{tt} =- g_{rr}^{-1}$, and the approximate solution is then obtained.  As a byproduct, we will show that the exact solutions can be extended to some typical limits which are related to Schwarzschild, Reissner-Nordstrom and Schwarzschild-de Sitter/anti-de Sitter solutions, respectively. We calculate the distribution function of the phantom field and its potential in the region of the galactic halo. In Sec. \ref{sec:def}, we will examine if the spatially inhomogeneous phantom field can stably survive with the existence of supermassive black hole at the center.
Considering the spherically symmetric space with a supermassive black hole at the center, we first semiclassically calculate the Klein-Gordon equation of  a $S$-wave phantom particle that is excited from the background field.  We obtain the absorption probability of the infalling phantom particle into the central black hole and further show that the absorptive cross section is approximately the area of the horizon. We show that the accretion rate of the phantom particle, which is accompanied by the decreasing rate of the black hole mass could be small enough, so that the space structure of a galaxy is stable compared with the life of the Universe.
In Sec. \ref{sec:stability}, we consider a toy model of  phantom particles coupled to massive dark matter, and show that both the dark matter  and {\it low frequency} phantom densities can be quasi-stable for a sufficiently small coupling constant.
Finally we give the summary in Sec. \ref{sec:sum}.

\section{The exact solution in the static limit}\label{sec:exact}

We consider the real phantom field minimally coupled to gravity
\begin{eqnarray}
S=\int d^4x \sqrt{-g} \left[ \frac{ R}{2 \kappa^2} +\frac{1}{2} g^{\mu\nu} \partial_\mu \Phi \partial_\nu \Phi -V(\Phi) +{\cal L}_m  +{\cal L}_I\right] \,,
\end{eqnarray}
where $\kappa^2=8\pi G$ is the reduced Planck mass, $V(\Phi)$ is the phantom field potential, the Lagrangian term ${\cal L}_m$ accounts for the massive dark matter in the galaxy, and ${\cal L}_I$ describes possible interactions between the phantom field and the dark matter.  Due to the small coupling between the phantom field and the dark matter, ${\cal L}_I$ can be negligible in the present calculation.  To investigate static, spherically symmetric solutions, we employ the metric
$ds^2 = -e^\nu dt^2 + e^{\lambda} dr^2 + r^2 (d\theta^2 + \sin^2\theta d\phi^2)$, adding the ansatz $\lambda=-\nu$. This condition is satisfied with the exact solutions, like the Schwarzschild, Reissner-Nordstrom and de Sitter/anti-de Sitter solutions. For the static situation, the Einstein equations read
 \begin{eqnarray}
&& g^{tt}R_{tt} - \frac{1}{2} R = e^{-\lambda} \left( \frac{1}{r^2} - \frac{\lambda^\prime}{r} \right) -\frac{1}{r^2} =\kappa^2 T^t_{\ t} \,, \label{eq:R-1}\\
&& g^{rr}R_{rr} - \frac{1}{2} R = e^{-\lambda} \left( \frac{1}{r^2} + \frac{\nu^\prime}{r} \right) -\frac{1}{r^2} = \kappa^2 T^r_{\ r} \,,  \label{eq:R-2}\\
&& g^{\theta\theta }R_{\theta\theta} - \frac{1}{2} R = \frac{e^{-\lambda}}{2} \left( \nu{''} +  \frac{\nu'^2}{2}   + \frac{\nu{'} -\lambda{'}}{r} - \frac{\nu^\prime \lambda^\prime}{2} \right)  = \kappa^2 T^\theta_{\ \theta}  \,, \label{eq:R-3}\\
&& g^{\phi\phi }R_{\phi\phi} - \frac{1}{2} R = \frac{e^{-\lambda}}{2} \left( \nu{''} +  \frac{\nu'^2}{2}   + \frac{\nu{'} -\lambda{'}}{r} - \frac{\nu^\prime \lambda^\prime}{2}  \right)  = \kappa^2 T^\phi_{\ \phi} \,.\label{eq:R-4}\end{eqnarray}
where the energy-momentum tensor corresponds to the massive dark matter in the background of the phantom field,
\begin{eqnarray}
  T^t_{\ t} &=& -\rho=-\rho_{ph} -\rho_{\rm DM} = \frac{1}{2} e^{-\lambda} \Phi^{\prime 2} -V(\Phi) -\rho_{\rm DM} \,, \label{eq:T-1}\\
   T^r_{\ r}    &=& p_r=p_{r, ph} =  -\frac{1}{2} e^{-\lambda} \Phi^{\prime 2} -V(\Phi) \,, \label{eq:T-2}\\
   T^\theta_{\ \theta}    &=& p_\theta=p_{\theta, ph} = \frac{1}{2} e^{-\lambda} \Phi^{\prime 2} -V(\Phi) \,,  \label{eq:T-3}\\
   T^\phi_{\ \phi}    &=& p_\phi=p_{\phi, ph} =  \frac{1}{2} e^{-\lambda} \Phi^{\prime 2} -V(\Phi) \,,  \label{eq:T-4}
   \end{eqnarray}
and a {\it prime} denotes  the differentiation with respect to $r$.
In the spherical coordinate, the total energy-momentum tensor is denoted by ${\rm diag}(-\rho, p_r, p_\theta, p_\phi)$, the phantom field background  is ${\rm diag}(-\rho_{ph}, p_{r,ph}, p_{\theta,ph}, p_{\phi,ph})$, and the cold dark matter can be approximated as ${\rm diag}(-\rho_{\rm DM}, 0,0,0)$.

It is interesting to note that we can find solutions that are satisfied with the condition $\lambda=-\nu$, {\it i.e.}  $g_{tt} = -g_{rr}^{-1}$,  in the limit $T^t_{\ t} \to  T^r_{\ r} $.  Moreover, it will be shown in the following that as for the mass density of WIMPs to be $\rho_{\rm DM}=e^{\nu} \Phi^{\prime 2}>0$, the present case can be satisfied with this condition for which  the existence of the corresponding solution is due to the fact that the sign of the kinetic term of the phantom field is opposite compared to the ordinary scalar field (quintessence field); for the quintessence dark energy, there exists no such solution.

To find the solution satisfied with the condition $\lambda=-\nu$,  we first set  $\nu= \ln (1-U)$ and substitute it into Eqs. (\ref{eq:R-1}) and (\ref{eq:R-3}), (or into  (\ref{eq:R-2}) and (\ref{eq:R-4})). We then obtain
 \begin{eqnarray}
&& r^2 U'' + 2 \epsilon\, r U' +2 (\epsilon-1) U=0 \,,
\end{eqnarray}
where we have set $T^\theta_{\ \theta}=T^\phi_{\ \phi}= T^t_{\ t}(1-\epsilon)$ with $\epsilon$ being a constant.
The solution of this equation is
 \begin{eqnarray}
 U=\frac{r_s}{r}  - \frac{r^{2(1-\epsilon)}}{r_\epsilon},  \ \  {\rm for}\  \epsilon \not= \frac{3}{2}  \,,
  \end{eqnarray}
or
\begin{eqnarray}
 U=\frac{1}{r}  \left[r_s -  a \ln \left( \frac{r}{|a|} \right) \right],  \ \  {\rm for}\  \epsilon = \frac{3}{2}  \,,
  \end{eqnarray}
i.e., the corresponding metric is
\begin{eqnarray}
ds^2 = - \left[ 1-\frac{r_s}{r}  + \frac{r^{2(1-\epsilon)}}{r_\epsilon}\right] dt^2  + \left[ 1-\frac{r_s}{r}  + \frac{r^{2(1-\epsilon)}}{r_\epsilon}\right]^{-1} dr^2
+r^2(d\theta^2 +\sin^2\theta d\phi^2) \,,  \label{eq:metric-1}
\end{eqnarray}
or
 \begin{eqnarray}
ds^2 = - \left[ 1-\frac{r_s}{r}  +  \frac{a}{r} \ln \left( \frac{r}{|a|} \right) \right] dt^2  + \left[ 1-\frac{r_s}{r}  +  \frac{a}{r} \ln \left( \frac{r}{|a|} \right) \right] ^{-1} dr^2
+r^2(d\theta^2 + \sin^2\theta d\phi^2) \,,
\end{eqnarray}
where $r_s, r_\epsilon$, and  $a\not=0$ are the integration constants.

Before we continue, we discuss the obtained exact solutions in some typical limits.  In addition to the present case of the galactic dark matter interacting with the phantom field, the following ones can be satisfied with the relation $T^r_r= T^t_{\ t}$. First, for $\epsilon=1$ with $r_\epsilon \to \infty$,  it gives the Schwarzschild metric corresponding to $T^{\mu}_{\ \nu}=0$, and $r_s$ is the Schwarzschild radius. Second, for $\epsilon=2$, the solution is the Reissner-Nordstrom metric, where $r_\epsilon^{-1} = G Q^2$ and $Q$ is the charge of the black hole.  Third, for $\epsilon=0$ resulting in  $ T^t_{\ t}= T^r_{\ r}=T^\theta_{\ \theta}=T^\phi_{\ \phi}= 3/(8\pi G r_\epsilon)$, it gives the Schwarzschild-de Sitter/anti-de Sitter solutions which are equivalent to the replacement $r_\epsilon \equiv -3/\Lambda$, with $\Lambda$ being the positive/negative cosmology constants.

In this paper, we are interested in the metric that can describe the motions of stars in the galaxy.  For the several observed cases, the rotational stars with radius $ r_{\rm halo}> r\gg  r_s$ in a spiral galaxy, where $r_{\rm halo}\gtrsim 50~{\rm kpc}$ denotes the radius of a typical halo in a galaxy, are moving in  circular orbits with {\it nearly} constant tangential speed $v$ which roughly ranges from $10^{-4}$ to $10^{-3}$.  For the region with $r>r_{\rm halo}$, the dark matter may become very dilute. In the dark matter dominant region, where the test particle stably moves in constant rotational curve, the  metric function $-g_{tt}$ was estimated in the form \cite{Matos:2000ki}
\begin{equation}\label{eq:constant_velocity}
-g_{tt} \cong\left( \frac{r}{r_0} \right)^{2v^2} \,, \label{eq:f_formC}
 \end{equation}
 where $r_0$ is a constant and $-g_{tt} =e^{\nu}$ will be denoted as $f$ in the following discussion. Note that the form of $g_{tt}$ is model independent but $g_{rr}$ is not \cite{Matos:2000ki,Boehmer:2007kx}. Because $2v^2\sim 10^{-8} - 10^{-6}$ is a tiny magnitude, we can approximate the metric function in the form
 \begin{eqnarray}
 f \cong \left( \frac{r}{r_0} \right)^{2v^2} = e^{\ln \left(\frac{r}{r_0}\right)^{2v^2}} \cong 1+ 2 v^2 \ln \frac{r}{r_0}\,. \label{eq:f_form1}
 \end{eqnarray}
 
\begin{figure}[t]
\begin{center}
\includegraphics[width=3.2in]{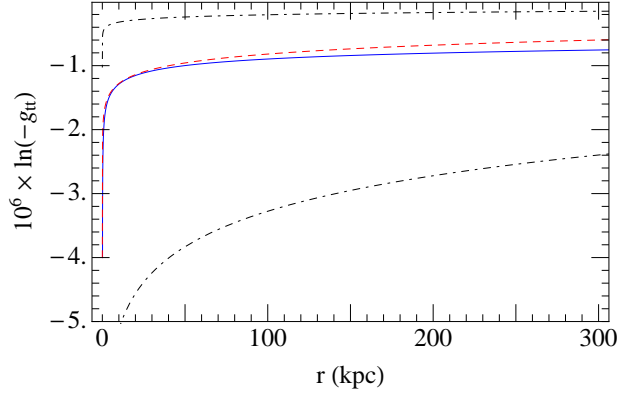}
\end{center}
\caption{The metric function $-g_{tt}$ as a function of $r$. The solid curve corresponds to $-g_{tt}=1- \frac{r_s}{r} +\gamma ( \frac{r}{\bar{r}_0} )^\alpha$ and the dashed curve is for $-g_{tt}=\left( \frac{r}{r_0} \right)^{2v^2}$, where we have adopted the following parameters: $r_s=10^{-7}$ pc, $r_0=6$ Mpc, $\alpha=-0.155$, $v^2=10^{-7}$. For comparison, using the same $r_0$, the upper and lower dot-dashed curves are for $-g_{tt}=\left( \frac{r}{r_0} \right)^{2v^2}$ but with $v^2=10^{-7}/4$ and  $v^2=4\times 10^{-7}$, respectively.}\label{fig:metric-function}
\end{figure}

On the other hand, we focus on our solution with $\epsilon \ll 1$ but $\not= 1$,  for which the galaxy has a supermassive black hole of the Schwarzschild radius $r_s\sim 10^{-7}\ {\rm pc}$ at its center and is predominated by the massive dark matter (WIMPs) in the phantom dark energy background. In the following, we will further exhibit that if we put a test particle in such a galaxy, it is possible to find a solution for which the test particle moves in circular orbit with nearly constant tangential speed consistent with the result given in Eq. (\ref{eq:f_form1})  (see the result given in Fig. \ref{fig:metric-function}). Compared with Eq. (\ref{eq:f_form1}), we find that in the same dark-matter dominant region  if the metric function in  Eq. (\ref{eq:metric-1}) is given by
  \begin{equation}
 f=1- \frac{r_s}{r} +\gamma \Bigg( \frac{r}{\bar{r}_0} \Bigg)^\alpha  \cong  1+\gamma e^{\ln \left(\frac{r}{\bar{r}_0}\right)^{\alpha}}
   \cong 1+ \gamma  \left(1+\alpha  \ln \frac{r}{\bar{r}_0} \right)\,,
\end{equation}
where $|\ln \left(\frac{r}{\bar{r}_0}\right)^{\alpha}|<1$, $ |\gamma|, |\alpha| \ll 1$,  $1-\epsilon=\alpha/2$, $r_\epsilon^{-1}=\gamma/\bar{r}_0^\alpha$, these two equations are approximately equal under the following conditions:  $\gamma\alpha =2v^2 \sim 10^{-8}-10^{-6}$ and $\gamma = 2v^2 \ln (\bar{r}_0 /r_0)$. In Fig. \ref{fig:metric-function}, we consider the extreme case of the dark-matter galaxy with the supermassive black hole at its center and show the corresponding metric function $-g_{tt}$ as a function of $r$. We find that the obtained metric function can be consistent with the result given in Eq. (\ref{eq:f_formC}) under the certain condition.

 In the region with the nearly constant tangential speed of the stars, the energy density is given by $\rho \simeq v^2 m_{\rm pl}^2/ 4\pi r^2$, where $m_{\rm pl}=G^{-1/2}$ is the Planck mass. Moreover, from Eqs. (\ref{eq:R-1}), (\ref{eq:R-2}), (\ref{eq:T-1}), and (\ref{eq:T-2}), we have $\rho_{\rm DM}=e^{\nu} \Phi_{b}^{\prime 2}$  in the static limit, where $\Phi_{b}$ is the classical phantom field background.  On the other hand, from Eqs. (\ref{eq:T-1}), (\ref{eq:T-2}) and (\ref{eq:T-3}) and using $\epsilon\approx 1$, we find that  $\rho = e^\nu \Phi_{b}'^2 /\epsilon \simeq \rho_{\rm DM} = e^\nu \Phi_{b}'^2$, i.e., $\rho_{ph}\simeq 0$ and $ V =  e^\nu \Phi_{b}'^2/2 +\rho (1-\epsilon) \simeq e^\nu \Phi_{b}'^2/2 $.  Since the metric function is  $e^{\nu} \simeq 1$, combining these gives

\begin{eqnarray}
\Phi_{b}(r)-\Phi_\infty &\approx& \frac{v m_{\rm pl}}{\sqrt{4\pi}} \ln \left( \frac{r}{\tilde{r}_0 }\right)\,, \\
V(\Phi_{b}) &\approx& \frac{v^2 m_{\rm pl}^2}{8\pi \tilde{r}_0^2  } e^{-\frac{\sqrt{16\pi}(\Phi_{b} -\Phi_\infty)}{v m_{\rm pl}}} \,, \label{eq:V}
\end{eqnarray}
during the distances $ r_{\rm halo}> r\gg  r_s$,   where the integration constant $\tilde{r}_0$ is roughly $\gtrsim 6$~Mpc, the  intergalactic distance. Here we choose the positive sign of $\Phi_b$. As for $r \gtrsim \tilde{r}_0$, the space becomes flat, the phantom field $\Phi _{b}=\Phi_\infty$ is spatially uniform, and its  potential  is responsible for the current accelerated expansion for which we have $3 H_0^2 \simeq 8\pi V(\Phi_\infty)/m_{\rm pl}^2  $, where at the present epoch the Hubble parameter $H_0\approx 10^{-42}$ GeV. Consistently, we get
\begin{eqnarray}
 \tilde{r}_0 \sim \frac{v}{H_0} \sim 6~ {\rm Mpc}.
\end{eqnarray}

 \section{The absorption of the phantom field by the supermassive black hole}\label{sec:def}

 The WIMPs (dark matter) are quite stable owing to the conservation of the angular momentum of the individual particles and the tiny interaction among them,  so that the capture cross section for dark matter particles by the supermassive black hole is sufficiently small.  Therefore, to study the stability of the space-time structure for the galaxy, we compute the accretion rate of  the excited phantom particles into the black hole. If the non-relativistic dark matter can be long lived enough compared with the age of the universe,
in the following, we will estimate the absorption probability, absorption cross section, and the accretion rate of the phantom  excitation wave into the black hole. At low energies, the dominant effect comes from the excited phantom particle with lowest angular momentum. Therefore, we consider its $S$ state, which is the excitation from the background, $\delta \Phi =\Phi(t,r) -\Phi_{b}(r)$.
The phantom potential is assumed to be around the local maximum,
\begin{eqnarray}
V(\Phi) &\approx& \frac{v^2 m_{\rm pl}^2}{8\pi \tilde{r}_0^2  } e^{-\frac{\sqrt{16\pi}(F(\Phi) -\Phi_\infty)}{v m_{\rm pl}}} \,,
\end{eqnarray}
where
$F(\Phi)\simeq \Phi_b + (F_{\Phi\Phi}(\Phi_b)/2) \delta\Phi^2$ (since the phantom particle with the negative kinetic energy might evolve to the maximum of the local potential). In other words, the local potential is around the locally stable point, $V_{\Phi}(\Phi_b)=0$. Here $F_{\Phi\Phi}\equiv d^2F/d\Phi^2$ and $V_{\Phi}\equiv dV/d\Phi$. Taking into account the back-reaction, the wave for the phantom excitation is satisfied with the Klein-Gordon equation
\begin{eqnarray}
  -\frac{1}{f} \partial_0^2 \delta\Phi + \frac{1}{r^2} \partial_r ( r^2 f \partial_r \delta\Phi) =  (m_\phi^2 +\lambda \langle \Psi^2 \rangle)\delta\Phi \equiv m_{\rm eff}^2 \delta\Phi \,,\label{eq: KG-1}
\end{eqnarray}
where
$$m_{\rm \phi}^2 =V(\Phi_b) \frac{\sqrt{16\pi}}{v m_{\rm pl}}  F_{\Phi\Phi}(\Phi_b)$$
 is the mass squared of the excited phantom field $\delta\Phi$, $\lambda$ is the dimensionless coupling constant describing the interaction between the excited phantom particle and the dark matter in the form of ${\cal L}_I= (\lambda/2) \delta\Phi^2 \cdot\Psi^2$, and $\langle \Psi^2 \rangle$ stands for the average of the quantum fluctuation of the  dark matter field $\Psi$. We will show in the next section that, when $\lambda$ is small enough, the densities $\rho_{\delta\Phi}$ and $\rho_\Psi$ exhibit a stable oscillatory behavior, so  that one can take $m_{\rm eff}^2 \simeq m_\phi^2$. In this paper, the conclusion is also applicable for $m_\phi^2<0$.

The excited phantom wave can be variable separated  in form of $\delta\Phi(t,r)=\Re[\phi(r) e^{-i \omega t}]$. In the present study, we reasonably assume the tiny phantom mass  and therefore neglect its spatial dependence.
 We redefine a new coordinate variable $\tau$ by the relation
 \begin{eqnarray}
 d\tau =\frac{dr}{r^2 f} \,.
 \end{eqnarray}
Then, the differential equation of $\phi$ reads
 \begin{equation}
 [\partial_\tau^2  + (\omega^2 - m_{\rm eff}^2 f) r^4] \phi =0 \,.
 \end{equation}
 Very close the horizon $r\approx r_s$, where $f\approx 0$ and we can have the approximation $(\omega^2 - m_{\rm eff}^2 f) r^4\simeq \omega^2 r_s^4$, the ingoing wave is then given by $\phi(r) \simeq \phi_0 e^{-i \omega r_s^2\tau}$. In this case, because the infalling phantom particles have a total negative energy, the phantom energy accretes resulting in the decrease of the black hole mass \cite{Babichev:2004yx}. Extending to the region,  where $r\gg \omega r_s^2$ and the rotational speed, $v$, of the star is constant, corresponding to $f\approx (r/r_0)^{2v^2}$ \cite{Matos:2000ki}, the solution can be approximated as
 \begin{eqnarray}
 \phi (r)\propto 1-i\omega  \left(1 -\frac{m_{\rm eff}^2}{\omega^2}\right)^{1\over 2} r_s^2 \tau =1+i d\omega r_s^2 \frac{r_0^{2v^2}}{r^{1+2v^2}} \frac{1}{1+2v^2} \,, \label{eq:asym-1}
 \end{eqnarray}
 where $d\simeq [1 -m_{\rm eff}^2/\omega^2]^{1/2} $.
 On the other hand,  using the dimensionless variable $\rho=\omega r$, we find  that the differential equation for $\phi$ can be rewritten by
\begin{eqnarray}
 \partial_\rho^2 \phi + \Bigg( \frac{ \partial_\rho f}{f} + \frac{2}{\rho} \Bigg) \partial_\rho \phi +\Bigg(\frac{1}{f^2} - \frac{m_{\rm eff}^2}{\omega^2 f} \Bigg)\phi =0 \,. \label{eq:EOM-2}
\end{eqnarray}
 For the distances $ r_{\rm halo}> r\gg  r_s$, this equation is approximated as
 \begin{eqnarray}
 \partial_\rho^2 \phi + \Bigg(  \frac{2v^2+2}{\rho} \Bigg) \partial_\rho \phi +\Bigg( 1-2\gamma- \frac{m_{\rm eff}^2}{\omega^2 }\Bigg)\phi =0 \,. \label{eq:EOM-2-approx}
\end{eqnarray}
The corresponding solution is
 \begin{equation}
 \phi(\rho) = \rho^{- ({1 \over 2} +v^2)}  \left(  A J_{{1 \over 2} +v^2}  (d\rho)+ B N_{{1 \over 2} +v^2}(d\rho)\right) \label{eq:solution-2}\,.
 \end{equation}
 The overlap region between the two solutions, Eqs. (\ref{eq:asym-1}) and (\ref{eq:solution-2}), is  $\omega^2 r_s^2 \ll \omega r \ll 1$, where Eq. (\ref{eq:solution-2}) reduces to
 \begin{eqnarray}
 \phi \simeq   \Bigg( \frac{d}{2} \Bigg)^{ {1\over 2}+v^2}  \frac{ A}{\Gamma( \frac{3}{2} +v^2)} - \Bigg( \frac{2}{d \omega^2} \Bigg)^{ {1\over 2}+v^2} \frac{\Gamma( \frac{1}{2} +v^2) B}{\pi r^{1+v^2}} \,. \label{eq:asym-2}
    \end{eqnarray}
Matching  Eq. (\ref{eq:asym-2}) onto Eq. (\ref{eq:asym-1}) in the overlap region, we find
 \begin{eqnarray}
  \frac{B}{A}=-i  \Bigg( \frac{d \omega}{2}  \Bigg)^{1+2v^2}
   \frac{d\omega r_s^2 \pi}{\Gamma( \frac{3}{2} +v^2)  \Gamma( \frac{1}{2} +v^2) r_0^{-2v} (1+2v^2)}
   =-i\delta  \approx -i d^2 \omega^2 r_s^2\,.
   \end{eqnarray}
From the above result, we expect $|B|\ll |A|$ for $\omega r_s \ll1$. For distances $\omega r\gg 1$, the solution in Eq.~(\ref{eq:solution-2}) asymptotically behaves as
\begin{eqnarray}
\phi \sim \frac{1}{\omega r \sqrt{2\pi d}} [(A-iB) e^{i\theta} + (A+iB) e^{-i\theta} ] \,, \label{eq:asym-phi-2}
\end{eqnarray}
  where $\theta=d \omega r - \pi v^2/2 -\pi/2$. Thus, we know that the absorption probability of a spherical $S$ wave is
\begin{eqnarray}
\Gamma=1-\Bigg|\frac{A-iB}{A+iB} \Bigg|^2 \cong \frac{4\delta}{(1+\delta)^2}\approx 4 d^2\omega^2 r_s^2\,.
\end{eqnarray}
 We then further calculate the absorption cross section for the ingoing $S$ wave.  For an incident plane wave $\phi(z) = e^{ikz}$, which can be expanded in the partial-wave amplitudes
\begin{eqnarray}
 e^{ikr\cos\theta} = \sum_\ell (2\ell +1) P_\ell (\cos\theta) \frac{e^{ikr}-  e^{-i(kr -\ell \pi)}}{2ikr}  ,
\end{eqnarray}
the absorption cross section of the $S$-wave component is
\begin{eqnarray}
\sigma_{\rm abs} &=&\frac{\rm number\ of \ particles\ absorbed\ by\ the\ area\ of\ spherical\ surface\ per\ unit \ time }{\rm number\ of \ incident \ particles\ crossing\ unit\ area\  per\ unit \ time} \nonumber\\
&=& \frac{\left| \frac{1}{2kr} \right|^2 4\pi  r^2\Gamma}{1}
 =4\pi r_s^2 \frac{d^2 \omega^2}{k^2} =4 \pi r_s^2,
\end{eqnarray}
where $k=d \omega$,  $r$ is the area of the spherical surface, and $4\pi r_s^2$ is exactly the  area of the Schwarzschild horizon. The result is consistent with the low-energy cross section for massless minimally coupled scalars \cite{Das:1996we}. It should be stressed that the result is independent of $r$. Because the phantom field has a negative kinetic term, the phantom energy flux onto the black hole is $T_{0r} = - \delta\Phi_{,t} \delta\Phi_{,r}$ which has the opposite sign compared with the ordinary matter fluid. For the ultralight $|m_\phi^2|$, the potential term can be negligible for $r< r_{\rm halo}$, and we can approximately take the Jacobson solution $\delta\Phi = \dot{\delta\Phi}_\infty [t + r_s \ln (1- r_s/r)]$ near the horizon \cite{Jacobson:1999vr}, where ${\delta\Phi}_\infty$ is the excitation of the phantom field in the absence of the black hole. The excited phantom particles carry negative energies.  As they fall into the black hole, the black hole diminishes with the rate $dM_{\rm BH}/dt = 4\pi r^2 T_{0 r} = -4 \pi r_s^2 (\dot{\delta\Phi}_\infty)^2$, accompanied by the accretion of the phantom energy \cite{Babichev:2004yx,LoraClavijo:2012vc,Gonzalez:2009jg}. In \cite{LoraClavijo:2012vc,Gonzalez:2009jg}, the full nonlinear absorption of a phantom field by a black hole has been taken into account. Alternatively, the above calculation can be obtained using the Eddington-Finkelstein coordinates for which the phantom energy flux across the absorption area (horizon) is $T_{vv}=-(\dot{\delta\Phi}_\infty)^2$, where the advance time coordinate $v=t+r+r_s\ln \frac{r-r_s}{r_s}$. Thus the decrease of the black hole mass with the same rate $dM_{\rm BH}/dt = 4 \pi r_s^2  T_{vv}=-4 \pi r_s^2 (\dot{\delta\Phi}_\infty)^2$.

The boundary condition far away the galaxy center is based on the assumption that the phantom field is spatially uniform. We adopt the  simple potential model $V(\Phi)=m^2 \Phi^2$ with $m\sim 10^{-33}$ eV \cite{Sami:2003xv} to describe the acceleration  phase at the present epoch.  We can  match this model with Eq. (\ref{eq:V}),  when $\Phi$ is very close to $\Phi_\infty \sim m_{\rm pl}$, If the kinetic energy of the phantom field tends to be subdominant compared with its potential energy, from the equation of motion we can have the approximation $\dot{\Phi}_\infty \simeq V_\Phi/(3H) = m m_{\rm pl}/\sqrt{6\pi}$. In this model no Big Rip occurs and it is satisfied with $w \to -1$ for $t \to \infty$. It is expected that the perturbations $\delta\Phi_\infty$ of the phantom field are of order $10^{-5} \Phi_\infty$ \cite{Carroll:2003st}. Therefore, if  the kinetic energy of the excited phantom  states is also expected to be much less than or even the same order of magnitude as that of the background, $(\dot{\delta\Phi}_\infty)^2 \sim  {\cal O} (\dot{\Phi}_\infty^2)$, the decrease rate of the black hole can be estimated to be $dM_{\rm BH}/dt \lesssim -10^{-21} M_\odot$ yr$^{-1}$, where we have used $M_{\rm BH}\approx 10^6 M_\odot$. Therefore, the decrease of the black hole mass is much less than a solar mass in the lifetime of the Universe.

\section{Stability} \label{sec:stability}

In this section, we  consider the classical evolution of a simple system for which a nonrelativistic  scalar dark matter couples to the excitation of the phantom field in a galaxy.  Note that to avoid the quantum instability of the vacuum at high frequencies, the phantom dark energy field defined in an effective theory is valid at low energy  \cite{Carroll:2003st,Cline:2003gs}.  The stringent limit for the momentum cutoff, which was estimated from the diffuse gamma ray background, is less than 3 MeV \cite{Cline:2003gs}. As for the phantom field defined in an effective theory at low energies, we will exhibit that this system can be quasi-stable for a weak coupling. It should be noted that even for low frequencies, the negative energy of phantom particles may cause the system to become unstable since the positive energy dark matter particles could increase the energy to any level as long as the phantom particles decrease the same magnitude of the energy. A similar system with the phantom excitation coupled to the massless graviton had been considered in Ref. \cite {Carroll:2003st}.

\begin{figure}[t]
\begin{center}
\includegraphics[width=3.2in]{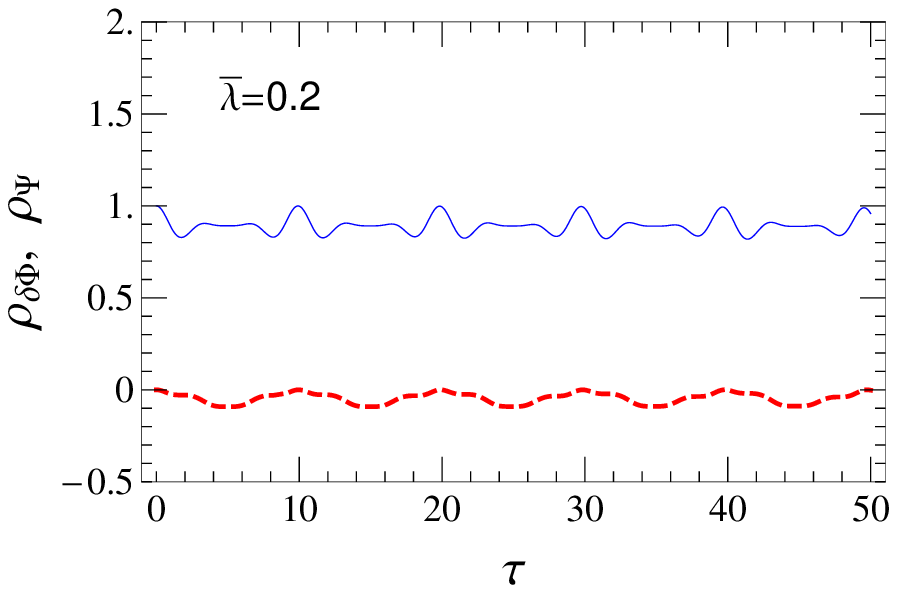}
\includegraphics[width=3.2in]{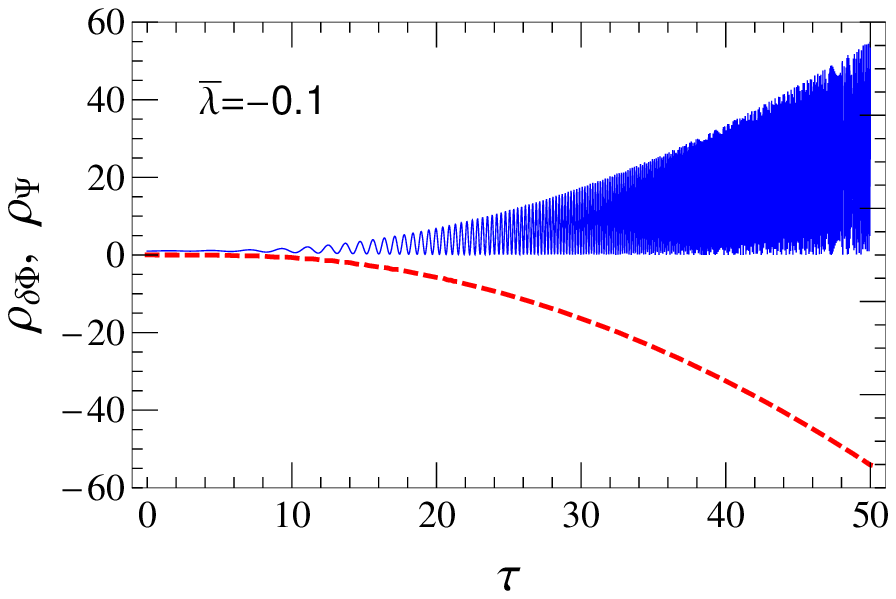}
\end{center}
\caption{Evolution of the energy densities of a coupled pair of the dark matter denoted by the solid curve and phantom excitation denoted by the dashed curve. The initial condition is $\bar{\Psi}'= \bar{\delta\Phi} '=0$ and $\bar{\Psi}= \bar{\delta\Phi} =1$ starting at $\tau=0$. The energy densities are displaced in units of $m_\Psi^2 M^2/2$ and satisfied with the constraint $\rho_\Psi + \rho_{\delta\Phi} +\rho_{\Psi\delta\Phi}=$ constant.}\label{fig:sho}
\end{figure}

The relevant Lagrangian ${\cal L}$  is
 \begin{eqnarray}
 {\cal L}=  -\frac{1}{2} g^{\mu\nu} \partial_\mu \Psi\partial_\nu \Psi  -\frac{1}{2} m_\Psi^2 \Psi^2
  + \frac{1}{2} g^{\mu\nu} \partial_\mu \delta\Phi\partial_\nu \delta\Phi + \frac{1}{2} m_\phi^2 \delta\Phi^2   +\frac{1}{2} \lambda \Psi^2  \delta\Phi^2\,,
  \end{eqnarray}
where $\Psi$ is the nonrelativisitic dark matter field and $\delta\Phi$ is the excitation of the phantom field as in the previous study. The spatial variations of them are relatively small in this consideration because $\Psi$ is nonrelativisitc and the accretion rate of the phantom excitations is typically small.  Thus the energy densities are
 \begin{eqnarray}
 \rho_\Psi &\simeq& \frac{1}{2} \dot{\Psi}^2 +\frac{1}{2} m_{\Psi}^2 \Psi^2 \,, \nonumber\\
 \rho_{\delta\Phi} &\simeq& -\frac{1}{2} \dot{\delta\Phi}^2 -\frac{1}{2} m_{\phi}^2 {\delta\Phi} ^2 \,,
  \end{eqnarray}
and the interaction term $\rho_{\Psi\delta\Phi}=- (\lambda/2) \Psi^2  \delta\Phi^2$. Note that here $\delta\Phi^2$ summarize only for low-energy modes. The equations of motion of these two kinds of particles are
 \begin{eqnarray}
 \bar{\Psi}'' & \simeq & - \left[  1- \bar{\lambda} \bar{\delta\Phi}^2     \right]  \bar{\Psi}\,, \nonumber\\
  \bar{\delta\Phi}'' & \simeq & - \left[  \frac{m_\phi^2}{m_\Psi^2}+ \bar{\lambda} \bar{\Psi}^2     \right]  \bar{\delta\Phi} \,,
  \end{eqnarray}
 where we have used the dimensionless variables
 \begin{eqnarray}
 \bar{\Psi} & =&\Psi/M \,,  \qquad   \bar{\delta\Phi}  =\delta\Phi/M  \,, \nonumber\\
  \bar{\lambda} & =&  \lambda (M/m_\Psi)^2\,, \quad \tau =m_\Psi t \,,
  \end{eqnarray}
 where $\rho_\Psi \simeq v^2 m_{\rm pl}^2/4\pi r^2$, and in the dark-matter dominant region of the galaxy, we have adopted the approximation for the metric function $f\simeq 1$. The perturbations, $\delta\Phi$ and $\Psi$, are expected to be of order $M\sim 10^{-5} m_{\rm pl}$ \cite{Carroll:2003st}. Note that in this section  a {\it prime} denotes  the differentiation with respect to $\tau$.
We plot the evolutions of densities in Fig. \ref{fig:sho}, where we have adopted the initial conditions, $\bar{\Psi}'= \bar{\delta\Phi} '=0$ and $\bar{\Psi}= \bar{\delta\Phi} =1$ at $\tau=0$. Although we have used $m_{\phi}=10^{-33}$ eV and $m_\Psi=10^{11}$ eV as inputs,  we note that evolution curves do not obviously change if the condition $m_{\phi}\ll m_\Psi$ is satisfied.
We obtain that for $0\leq\bar\lambda \leq 0.37$, the densities oscillate with a stable behavior, where $\rho_{\delta\Phi}\sim 0$. However, for  $\bar\lambda<0$,  $\rho_\Psi$ and $-\rho_{\delta\Phi}$ grow quickly.   In short, we could conclude that the stability for the galactic cold dark matter existing in the phantom field background is possible.
%
%


\section{Summary}\label{sec:sum}

 We have studied the possibility that the galactic dark matter exists in an scenario where the phantom field, responsible for the dark energy, may not be spatially homogeneous in a galaxy. We have obtained the statically and spherically approximate solution for this kind of galaxy system with a  supermassive black hole at its center.  The static exact solution of the metric functions is satisfied with $g_{tt} = - g_{rr}^{-1}$,  of which the relation is also consistent with the black hole solutions in the vacuum, electromagnetic and cosmological constant sources, corresponding to Schwarzschild, Reissner-Nordstrom and Schwarzschild-de Sitter/anti-de Sitter metrics, respectively.

Constrained by the observation that  the rotational stars in a spiral galaxy are moving in  circular orbits with nearly constant tangential speed, we have obtained that the background of the phantom field which is spatially inhomogeneous has (i) the small density $\rho_{ph}\simeq 0$ with negligible fluctuation if the coupling between the excitation phantom field and dark matter is small, and (ii) an exponential potential.

To avoid the well-known quantum instability of the vacuum at high frequencies, the phantom field that we consider here is an effective theory valid at low energies. Under this condition, we have computed the absorption cross section of the $S$-wave excitations, arising from the phantom background, into the central black hole and shown that it is equal to $4\pi r_s^2$,  the horizontal area of the central black hole.  Because the infalling phantom particles have a total negative energy, we estimate that the black hole mass thus diminishes at the quite small rate $dM_{\rm BH}/dt \lesssim -10^{-21} M_\odot$ yr$^{-1}$, so that  the decrease of the black hole mass is much less than a solar mass in the lifetime of the Universe. Furthermore, using a simple model with the cold dark matter very weakly coupled to the "low-frequency" phantom particles which are generated from the background,  we have shown that the dark matter and phantom densities can be quasi-stable.

\section*{Acknowledgments}

We are grateful to Dr. Ho-Chin Tsai for useful discussions.
This research was supported in part by the National Center for Theoretical Sciences and the
National Science Council of R.O.C. under Grant No. NSC99-2112-M-033-005-MY3.



\begin{thebibliography}{99}

\bibitem{Perlmutter:1998np}
  S.~Perlmutter {\it et al.}  [Supernova Cosmology Project Collaboration],
  Astrophys.\ J.\  {\bf 517}, 565 (1999)
  [arXiv:astro-ph/9812133].

\bibitem{Riess}
A. G. Riess {\it et al.}, Astron. J. 116,1009 (1998); Astron. J. 117,707 (1999).


\bibitem{Komatsu:2010fb}
  E.~Komatsu {\it et al.}  [WMAP Collaboration],
  Astrophys.\ J.\ Suppl.\  {\bf 192}, 18 (2011)
  [arXiv:1001.4538 [astro-ph.CO]].



\bibitem{Alam:2003fg}
  U.~Alam, V.~Sahni, T.~D.~Saini and A.~A.~Starobinsky,
  Mon.\ Not.\ Roy.\ Astron.\ Soc.\  {\bf 354} (2004) 275
  [arXiv:astro-ph/0311364].

\bibitem{Caldwell:1999ew}
  R.~R.~Caldwell,
  Phys.\ Lett.\  B {\bf 545}, 23 (2002)
  [arXiv:astro-ph/9908168].

\bibitem{Caldwell:2003vq}
  R.~R.~Caldwell, M.~Kamionkowski and N.~N.~Weinberg,
  Phys.\ Rev.\ Lett.\  {\bf 91}, 071301 (2003)
  [arXiv:astro-ph/0302506].

\bibitem{Cai:2009zp} 
  Y.~-F.~Cai, E.~N.~Saridakis, M.~R.~Setare and J.~-Q.~Xia,
  Phys.\ Rept.\  {\bf 493}, 1 (2010)
  [arXiv:0909.2776 [hep-th]].

\bibitem{Nesseris:2004uj}
  S.~Nesseris and L.~Perivolaropoulos,
  Phys.\ Rev.\  D {\bf 70}, 123529 (2004)
  [arXiv:astro-ph/0410309].

  \bibitem{spergel:2003}
D. N. Spergel et al., Astrophys. J. Suppl. {\bf 148}, 175
(2003).

\bibitem{Carroll:2003st}
  S.~M.~Carroll, M.~Hoffman and M.~Trodden,
  Phys.\ Rev.\  D {\bf 68}, 023509 (2003)
  [arXiv:astro-ph/0301273].


\bibitem{Copeland:2006wr}
  E.~J.~Copeland, M.~Sami and S.~Tsujikawa,
  Int.\ J.\ Mod.\ Phys.\  D {\bf 15}, 1753 (2006)
  [arXiv:hep-th/0603057].

\bibitem{Guo:2004xx}
  Z.~K.~Guo, R.~G.~Cai and Y.~Z.~Zhang,
  JCAP {\bf 0505}, 002 (2005)
  [arXiv:astro-ph/0412624].

\bibitem{Sushkov:2005kj}
  S.~V.~Sushkov,
Phys.\ Rev.\ D {\bf 71}, 043520 (2005)  [gr-qc/0502084].  

\bibitem{Zaslavskii:2005fs}
  O.~B.~Zaslavskii,
Phys.\ Rev.\ D {\bf 72}, 061303 (2005)  [gr-qc/0508057].  

\bibitem{Kuhfittig:2006xj}
  P.~K.~F.~Kuhfittig,
Class.\ Quant.\ Grav.\  {\bf 23}, 5853 (2006)  [gr-qc/0608055].  

\bibitem{Lobo:2005us}
  F.~S.~N.~Lobo,
Phys.\ Rev.\ D {\bf 71}, 084011 (2005)  [gr-qc/0502099].  

\bibitem{Lobo:2005yv}
  F.~S.~N.~Lobo,
Phys.\ Rev.\ D {\bf 71}, 124022 (2005)  [gr-qc/0506001].  

\bibitem{Lobo:2005uf}
  F.~S.~N.~Lobo,
Class.\ Quant.\ Grav.\  {\bf 23}, 1525 (2006)  [gr-qc/0508115].  

\bibitem{DeBenedictis:2008qm}
  A.~DeBenedictis, R.~Garattini and F.~S.~N.~Lobo,
Phys.\ Rev.\ D {\bf 78}, 104003 (2008)  [arXiv:0808.0839 [gr-qc]].  

\bibitem{Dzhunushaliev:2008bq}
  V.~Dzhunushaliev, V.~Folomeev, R.~Myrzakulov and D.~Singleton,
JHEP {\bf 0807}, 094 (2008)  [arXiv:0805.3211 [gr-qc]].  

\bibitem{Yazadjiev:2011sm}
  S.~S.~Yazadjiev,
Phys.\ Rev.\ D {\bf 83}, 127501 (2011)  [arXiv:1104.1865 [gr-qc]].  

\bibitem{Salgado:2003ub}
  M.~Salgado,
  Class.\ Quant.\ Grav.\  {\bf 20}, 4551 (2003)
  [arXiv:gr-qc/0304010].

\bibitem{Dymnikova:2001fb}
  I.~Dymnikova,
  Class.\ Quant.\ Grav.\  {\bf 19}, 725 (2002)
  [arXiv:gr-qc/0112052].

\bibitem{Giambo:2002wr}
  R.~Giambo,
  Class.\ Quant.\ Grav.\  {\bf 19}, 4399 (2002)
  [arXiv:gr-qc/0204076].

\bibitem{Kiselev:2002dx}
  V.~V.~Kiselev,
  Class.\ Quant.\ Grav.\  {\bf 20}, 1187 (2003)
  [arXiv:gr-qc/0210040].
  
\bibitem{Cline:2003gs}
  J.~M.~Cline, S.~Jeon and G.~D.~Moore,
  Phys.\ Rev.\ D {\bf 70}, 043543 (2004)
  [hep-ph/0311312].

\bibitem{Matos:2000ki}
  T.~Matos, F.~S.~Guzman and D.~Nunez,
  Phys.\ Rev.\  D {\bf 62}, 061301 (2000)
  [arXiv:astro-ph/0003398].

\bibitem{Boehmer:2007kx} 
  C.~G.~Boehmer, T.~Harko and F.~S.~N.~Lobo,
  Astropart.\ Phys.\  {\bf 29}, 386 (2008)
  [arXiv:0709.0046 [gr-qc]].
  
\bibitem{Babichev:2004yx}
  E.~Babichev, V.~Dokuchaev and Yu.~Eroshenko,
  Phys.\ Rev.\ Lett.\  {\bf 93}, 021102 (2004)
  [arXiv:gr-qc/0402089].

\bibitem{Das:1996we}
  S.~R.~Das, G.~W.~Gibbons and S.~D.~Mathur,
  Phys.\ Rev.\ Lett.\  {\bf 78}, 417 (1997)
  [arXiv:hep-th/9609052].

   
\bibitem{Jacobson:1999vr}
  T.~Jacobson,
  Phys.\ Rev.\ Lett.\  {\bf 83}, 2699 (1999)
  [arXiv:astro-ph/9905303].
  
\bibitem{LoraClavijo:2012vc} 
  F.~D.~Lora-Clavijo, J.~A.~Gonzalez and F.~S.~Guzman,
  AIP Conf.\ Proc.\  {\bf 1256}, 339 (2010)
  [arXiv:1207.3375 [astro-ph.CO]].
 
\bibitem{Gonzalez:2009jg} 
  J.~A.~Gonzalez and F.~S.~Guzman,
  Phys.\ Rev.\ D {\bf 79}, 121501 (2009)
  [arXiv:0903.0881 [gr-qc]].



\bibitem{Sami:2003xv}
  M.~Sami and A.~Toporensky,
  Mod.\ Phys.\ Lett.\  A {\bf 19}, 1509 (2004)
  [arXiv:gr-qc/0312009].

\end{thebibliography}
 \end{document}